\documentclass[aip,reprint,showkeys,longbibliography]{revtex4-2}

\usepackage[colorlinks=true, citecolor=blue, urlcolor=blue, linkcolor=black]{hyperref}

\makeatletter
\let\@fnsymbol\@fnsymbol@latex
\@booleanfalse\altaffilletter@sw
\makeatother

\usepackage{graphicx}
\usepackage{float}
\usepackage{mathptmx}
\usepackage{amsmath}
\usepackage{latexsym}
\usepackage{color,colortbl}
\usepackage[table]{xcolor}
\usepackage{tabularx}
\usepackage{hyperref}
\usepackage{float}
\usepackage{footmisc}

\begin{document}
\title{Evolution of the public opinion on COVID-19 vaccination in Japan}

\date{\today}

\author{Yuri Nakayama}
\thanks{These authors contributed equally.}
\affiliation{Graduate School of Frontier Sciences, The University of Tokyo, Chiba, Japan}

\author{Yuka Takedomi}
\thanks{These authors contributed equally.}
\affiliation{National Institute of Informatics, Tokyo, Japan}

\author{Towa Suda}
\thanks{These authors contributed equally.}
\affiliation{National Institute of Informatics, Tokyo, Japan}

\author{Takeaki Uno}
\affiliation{National Institute of Informatics, Tokyo, Japan}

\author{Takako Hashimoto}
\affiliation{Faculty of Commerce and Economics, Chiba University of Commerce, Chiba, Japan}
\affiliation{Institute of Industrial Science, The University of Tokyo, Tokyo, Japan}

\author{Masashi Toyoda}
\affiliation{Institute of Industrial Science, The University of Tokyo, Tokyo, Japan}

\author{Naoki Yoshinaga}
\affiliation{Institute of Industrial Science, The University of Tokyo, Tokyo, Japan} 

\author{Masaru Kitsuregawa}
\affiliation{Institute of Industrial Science, The University of Tokyo, Tokyo, Japan}
\affiliation{National Institute of Informatics, Tokyo, Japan}

\author{Luis E.C. Rocha}
\email{luis.rocha@ugent.be}
\affiliation{Department of Economics, Ghent University, Ghent, Belgium}
\affiliation{Department of Physics and Astronomy, Ghent University, Ghent, Belgium}

\author{Ryota Kobayashi}
\email{r-koba@k.u-tokyo.ac.jp}
\thanks{This author contributed equally to Yuri Nakayama, Yuka Takedomi, and Towa Suda.} 
\affiliation{Graduate School of Frontier Sciences, The University of Tokyo, Chiba, Japan}
\affiliation{Mathematics and Informatics Center, The University of Tokyo, Tokyo, Japan}
\affiliation{JST PRESTO, Saitama, Japan}

\begin{abstract}
Vaccines are promising tools to control the spread of COVID-19. 
An effective vaccination campaign requires government policies and community engagement, sharing experiences for social support, and voicing concerns to vaccine safety and efficiency. 
The increasing use of online social platforms allows us to trace large-scale communication and infer public opinion in real-time. We collected more than 100 million vaccine-related tweets posted by 8 million users and used the Latent Dirichlet Allocation model to perform automated topic modeling of tweet texts during the vaccination campaign in Japan. We identified 15 topics grouped into 4 themes on Personal issue, Breaking news, Politics, and Conspiracy and humour. The evolution of the popularity of themes revealed a shift in public opinion, initially sharing the attention over personal issues (individual aspect), collecting information from the news (knowledge acquisition), and government criticisms, towards personal experiences once confidence in the vaccination campaign was established. An interrupted time series regression analysis showed that the Tokyo Olympic Games affected public opinion more than other critical events but not the course of the vaccination. Public opinion on politics was significantly affected by various events, positively shifting the attention in the early stages of the vaccination campaign and negatively later. Tweets about personal issues were mostly retweeted when the vaccination reached the younger population. The associations between the vaccination campaign stages and tweet themes suggest that the public engagement in the social platform contributed to speedup vaccine uptake by reducing anxiety via social learning and support.
\end{abstract}

\maketitle

\section{INTRODUCTION}
Vaccination is an effective mechanism to reduce the number of hospitalisations and mortality caused by the emergent coronavirus disease (COVID-19). With the advent of efficient vaccines after the first wave of the COVID-19 pandemics, public health efforts turned on strategies to cost-effectively immunise the population to increase survival and return economic activity.
The availability of doses and uptake rates are fundamental aspects to reach a sufficient vaccination coverage but those numbers varied across countries in the current pandemics. One particular concern was the hesitancy against the safety and effectiveness of COVID-19 vaccines~\cite{Wouters2021}, that affected the individual willingness to get vaccinated not only in low- and middle-income~\cite{Lazarus2021,Solis2021} but also in high-income countries~\cite{Dror2020,Fisher2020, Neumann2020,Nomura2021}. Japan stood out among developed economies as having one of the lowest vaccine confidence levels in the population~\cite{DeFigueiredo2020}. This results from safety concerns about the Human papillomavirus (HPV) vaccine that emerged in the early 2010s as a result of misinformation spread on adverse effects of the HPV vaccine~\cite{Hanley2015,Simms2020}, prompting the Japanese Ministry of Health, Labour and Welfare to suspend the proactive recommendations of HPV vaccine from June 2013 until November 2021. Such low public confidence pushed back the start of mass vaccination against COVID-19 for two months behind the USA, China, and European countries, leading to safety concerns and inquiries on the Tokyo Olympic Games that had already been postponed to August 2021. Albeit late, Japan achieved high vaccination coverage in a short time, becoming one of the highest in the world (ranked 13th among 229 countries) and reaching $77.8\%$ (at least one dose) on the 31st of October 2021~\cite{Mathieu2021}, ranking ahead of early adopters such as the UK ($73.3\%$), Germany ($69.0\%$), and the USA ($66.7\%$).

It is unclear how public opinion affected government policies that were also pressed by the domestic economic slowdown and the concerns about the Tokyo Olympic Games. On the other hand, public opinion typically reacts to policies and might serve as a barometer of government strategies. Monitoring public opinion however is challenging. The largest study of vaccination intention in Japan surveyed 30 000 participants~\cite{Nomura2021} and found that a large proportion of the population was unsure ($33\%$) or unwilling ($11\%$) to take COVID-19 vaccine, with side effects and safety being the main reasons. Classic survey studies like this are costly, relatively slow, and with few exceptions~\cite{DeFigueiredo2020} cannot trace changes of public opinion in real-time~\cite{Neumann2020,Lazarus2021,Murphy2021,Nomura2021}. Large-scale studies aiming to increase accuracy and the spatio-temporal resolution of responses require advanced survey techniques. In recent years, Human activity has been increasingly mediated by digital devices, leaving footprints that can be exploited to assess the population health and opinions~\cite{Chew2010,Aramaki2011, Proskurnia2017, Sinnenberg2017,Du2021}. 
In the context of COVID-19, social media data have been used to predict the number of new cases (incidence)~\cite{O2020,Qin2020} and to interpret the public perception of the pandemics~\cite{Tsao2021}. 
Twitter has been particularly useful to monitor public opinion because users engage and react timely to environmental changes, as for example reacting
to epidemic outbreaks~\cite{Medford2020,Boon2020}, expressing concerns on the disease~\cite{Abd2020}, 
accepting the pandemic situation~\cite{Aiello2021}, or on vaccination issues~\cite{Lyu2021,Kwok2021,Wu2021}. 

We hypothesise that Twitter activity can be used as a barometer of public concerns and response to government policies on the mass vaccination campaign in Japan during the COVID-19 pandemics. Twitter is widely used in Japan, where more than $60\%$ of the population below 40 years old is actively engaged~\cite{Soumu2021}. The pervasiveness of Twitter provides a unique source of data to monitor the evolution of the public opinion during the various stages of the Japanese vaccination campaign. To validate our hypothesis, we monitored over 8 million users (approx. $6.4\%$ of the Japanese population) and collected over 100 million tweets (including 76 million retweets) written in Japanese from January 1 to October 31, 2021. Our sample covers the period before the start of the vaccination campaign until weeks after the end of the Olympic Games in Tokyo. We perform topic modelling~\cite{Mills2014, Salmons2016} by applying a methodology based on the Latent Dirichlet Allocation (LDA) model~\cite{Blei2003} to disentangle the textual information and find underlying semantic structures in tweets.

\section{Results}

\begin{figure}[t]
  \centering
       \includegraphics[width=1.0\linewidth]{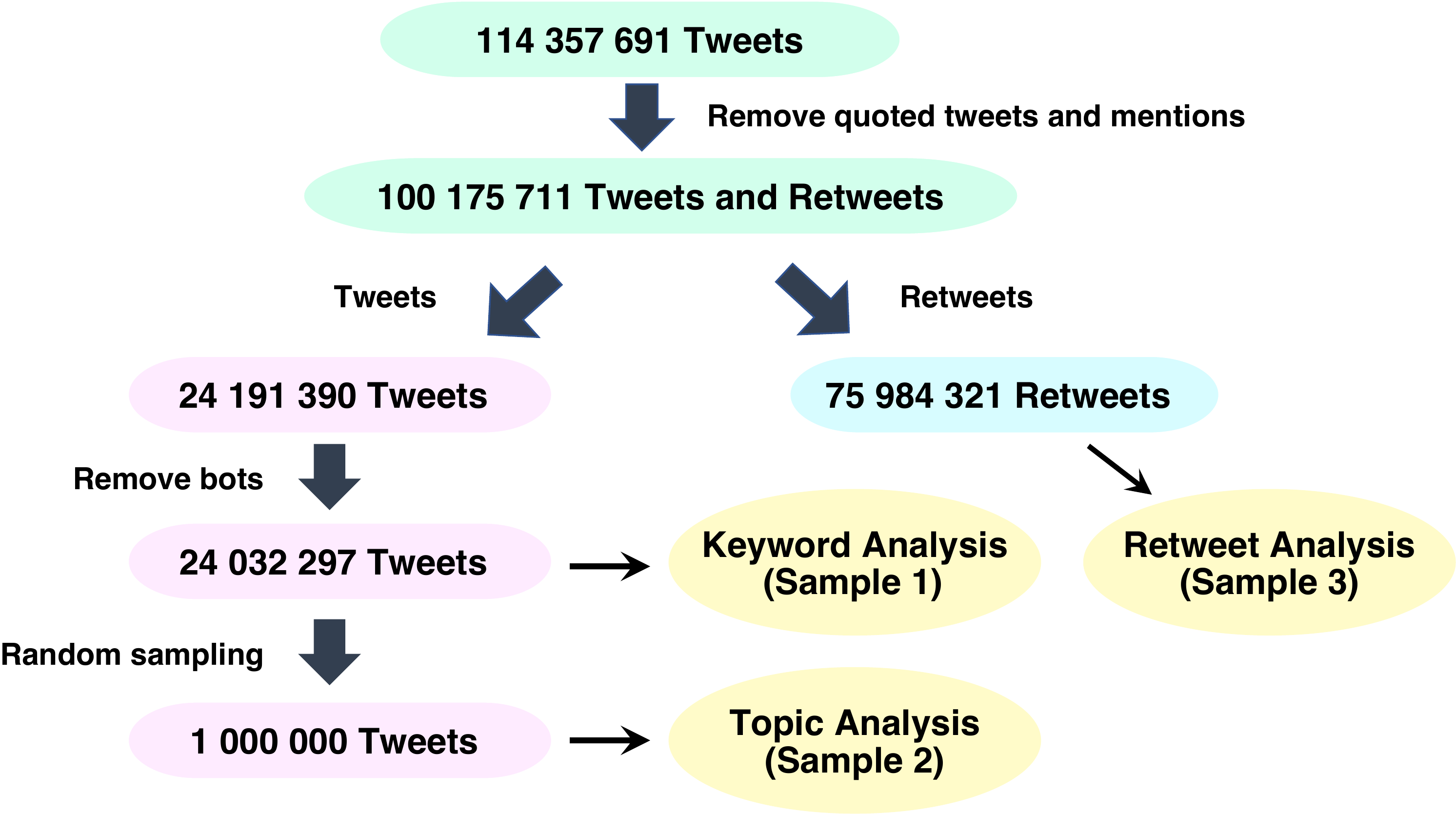}
  \caption{ {\bf  Data processing workflow.} 
}
  \label{fig:Data_Processing}
\end{figure}

The original data set contains 114 357 691 vaccine-related tweets written in Japanese from January 1 to October 31, 2021. Our analysis is based on three samples containing either the original tweets or retweets 
The first sample (sample 1) contains 24 032 297 tweets posted by
6 034 434 users 
and is used to study the evolution of the opinions, including disruptions due to critical events. A random sample of the original data (sample 2) is then used to identify the main topics and themes, and a sample of all retweets (sample 3) is used to study the spread of opinions~(Fig.~\ref{fig:Data_Processing}). 


\begin{figure}[thb]
  \centering
       \includegraphics[width=1.0\linewidth]{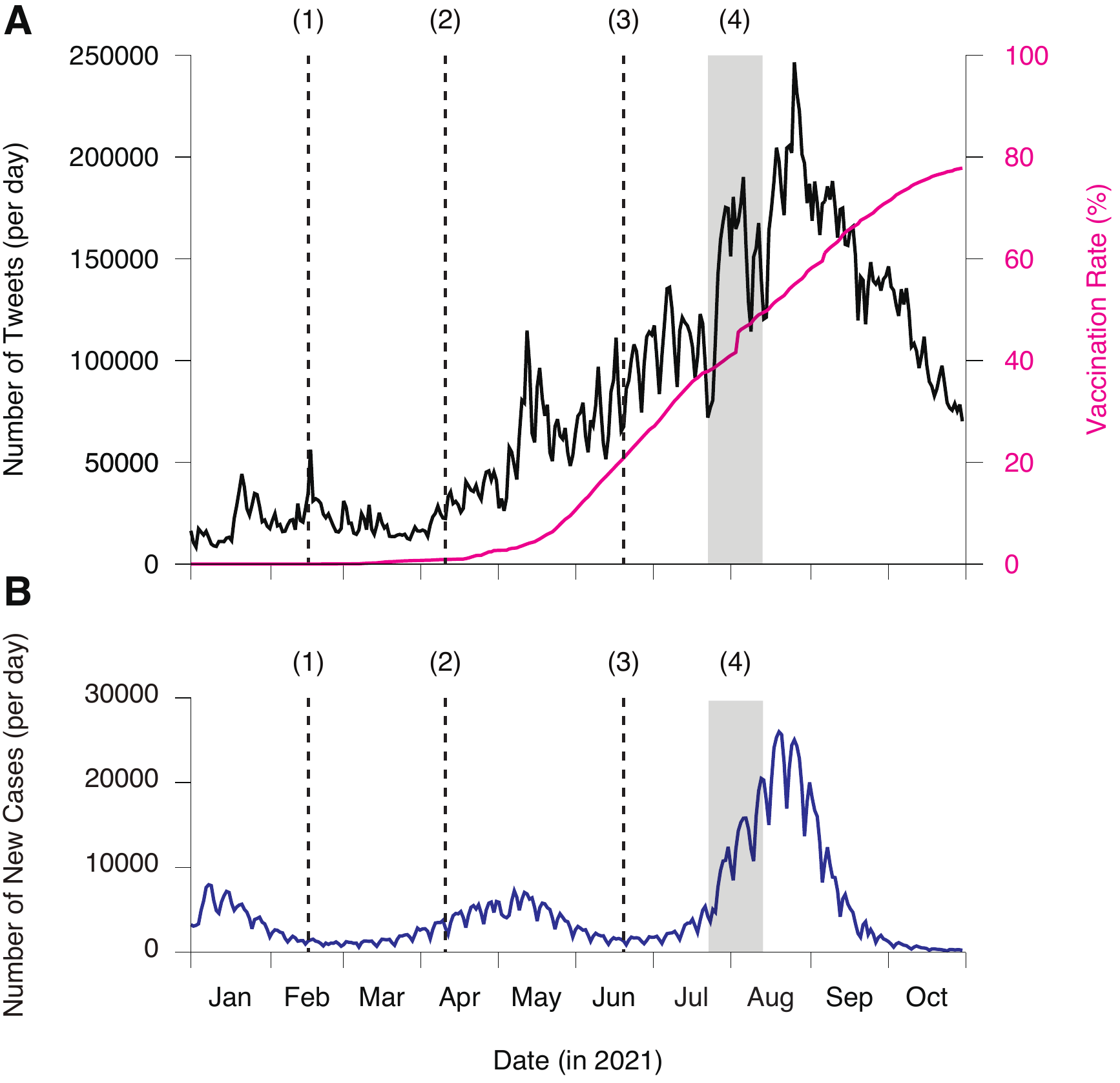}
  \caption{ {\bf  Vaccine related tweets, vaccination rate, and incidence of COVID-19.} (A) Number of vaccine-related tweets per day written in Japanese (black, left y-axis) and fraction of the vaccinated population in Japan (magenta, right y-axis) between January 1 and October 31, 2021. (B) Daily incidence of COVID-19 in Japan. Vertical lines indicate four main events during the study period: 1) launch of the COVID-19 vaccination campaign for essential workers; 2) for the elderly population (above 65 years old); and 3) for the general population (under 65 years old); 4) Period of the Tokyo Olympic Games.}
  \label{fig:Twitter_TS}
\end{figure}

\subsection{Vaccine Related Tweets}
\label{sec:Freq_Tweet}

Figure~\ref{fig:Twitter_TS}A shows the number of (vaccine-related) tweets per day during the study period and highlights four critical events~\cite{Covid19_Summary_JP}: 1) the launch of the COVID-19 vaccination campaign by the Japanese government, on February 17, focusing initially on essential workers (e.g. healthcare workers); 2) the start of vaccination of the elderly population (above 65 years old) on April 12; 3) the start of the general public vaccination on June 21; and 4) the Tokyo Olympic Games taking place from July 23 to August 8. The first peak occurs on January 21, when the Prime Minister Yoshihide Suga made an statement that ``the high coverage of the vaccination is not a precondition for holding the Olympics in Tokyo'' and the Ministry of Health, Labour and Welfare (MHLW) signed a contract with Pfizer Inc. to supply a total of 72 million doses of its COVID-19 vaccine. Although a spike is observed at the very start of the vaccination campaign (event 1), vaccine-related tweets started to increased after event 2, when the vaccination of non-essential workers began. That coincides with the outbreak of the 4th wave in Japan (early April, Fig.~\ref{fig:Twitter_TS}B) and is followed by increased interest during the peak of infections in the 4th wave (May 13), when the online booking of vaccine appointments was launched but became overwhelmed, leaving many people without a vaccination slot. There was a sharp relative decrease in the tweets at the start and end of the Olympic Games, followed by the largest peak on August 26, when a contamination scandal (approx. 1.6 million doses of the Moderna vaccine were discarded) was publicized. This last peak also coincides with the peak of the 5th wave and is followed by a substantial decrease on vaccine-related tweets, likely because of the high vaccination rate in the population~(Fig.~\ref{fig:Twitter_TS}A,B). 

\subsection{Clustering Vaccine Related Tweets}

The ranking of the most used words on vaccine-related tweets (sample 2) reveals that 242 627 ($24\%$) of them explicitly contained the word ``COVID-19'' (SI Appendix, Table~S1). While the prevalence of specific words in tweets can reveal patterns of popular words, this measure is unable to unveil hidden semantic relations among the tweets. We thus apply a machine learning methodology, the Latent Dirichlet Allocation (LDA) model, on a sample of 1 000 000 tweets (100 000 per month, sample 2) to automatically identify and classify (i.e.\ cluster) tweets into meaningful topics. This monthly sampling is used to remove the non-stationarity of the tweet activity given the unbalance in the number of vaccine-related tweets during the study period. 
Using LDA, we automatically identified 15 topics from the tweets (solely based on the textual content) and manually grouped them into four general themes: 1) Personal issue, 2) Breaking news, 3) Politics, and 4) Conspiracy and humour.
Table~\ref{tab_topic1} shows examples of representative tweets in each theme and Table~S2 (SI Appendix) shows the most popular words in each topic.

The most popular theme emerging from the topic analysis was Personal issue ($49.8\%$) and is formed by two topics about personal issues before being vaccinated, i.e.\ personal view on the vaccination and personal schedule of vaccination, and four topics about personal experience after being vaccinated, i.e.\ a topic about live reporting on the vaccination experience (e.g.\ waiting room or to/from the vaccination center), and three topics about individual vaccination experience including: 1) complaints about discomfort, and side effects and personal life after the vaccination; 
2) reporting body temperature after taking the vaccine; and 3) advice to overcome side effects  (Table~\ref{tab_topic1}).


The second most popular theme was classified as Breaking news ($21.3\%$), that includes two topics about news on COVID-19 vaccine such as vaccine development and approval, and the vaccine effectiveness. The first topic includes tweets about the development of Moderna, AstraZeneca, and Pfizer vaccines (clinical trials and government approval) in Japan and other countries. The second topic is about the effectiveness of vaccines and contains information about mRNA vaccines, the effectiveness of vaccines against new variants, and serious side effects (e.g.\ trombus) of the AstraZeneca vaccine. The last topic is about booking an appointment for vaccination,
in particular about the availability and whether users could successfully book a timeslot or not (Table~\ref{tab_topic1}).

Politics was classified as the third most popular theme ($17.2\%$) with three topics. The first topic was related to opinions about the government. For instance, the users complained that the vaccination schedule in Japan was behind other countries and disagreed on holding the Tokyo Olympic Games given the low vaccination coverage. Opinions about the mass media, such as the complaints about the unreliable information from the media and the attitude of the press inciting unrest, formed the second topic. Finally, the vaccination policy, including casual chats, e.g.\ tweets mentioning the assignment of Mr Taro Kono (a politician famous among the young population) as vaccine Minister, formed the third topic (Table~\ref{tab_topic1}).  
 
The least popular theme contained topics related to Conspiracy and humour ($11.8\%$). The first topic is about the control of the population, as for example the conspiracy theory that ``the purpose of COVID-19 vaccination was to reduce the global population'', whereas the second topic was about effects on the body, as for example that ``COVID-19 vaccines are a ploy to connect people to the 5G network''. Internet memes formed the third topic, as for example the popular ``Vac-vac-cine-cine'' (from ``vaccine vaccine'' because a person needs two vaccine shots to be fully vaccinated and because the combination of these words sounds like ``exciting'' and ``male genitalia'' in Japanese) (Table~\ref{tab_topic1}).

\begin{table*}[t]	\centering
\caption{Topics automatically identified on vaccine-related tweets before and during the COVID-19 vaccination campaign in Japan.  \label{tab_topic1} }
\vspace{3mm}
\tabcolsep = 10pt
\begin{tabular}{p{5.5cm}p{2cm}p{8.5cm}}
 \hline
 Themes and Topics 	& No. of tweets ($\%$) 	& Representative tweet (original tweets are in Japanese)   \\
 \hline\hline
{\bf Theme 1: Personal issue} 	                    & {\bf $49.8$}      &   	\\
Personal view on vaccination 		                    & $17.2$ 	& 
\textit{``I'll be vaccinated because I want to. If you don't want to, you don't have to. I don't think I should tell others to be vaccinated!''} \\
Personal schedule of vaccination 		                & $5.8$ 	& 
\textit{``I'm finally getting the Pfizer covid-19 vaccine tomorrow. I’m so excited.''}  \\
Live reports of before/after vaccination 		    & $3.2$ 	&  
\textit{``I've arrived the vaccination venue too early. I'm waiting and killing time:)''}  \\
Journal about vaccination experience 		            & $13.4$ 	&   \textit{``I've got the 2nd shot. I was fine after the 1st shot, but I don't think I'm fine this time. Side effect will come sooner or later.''}   \\
Perception after vaccination 		        & $6.6$ 	&   
\textit{``The injection was given very quickly and was not very painful. It has been five hours since the injection, and I feel a little bit of discomfort in my left arm...''}   \\
Preparation for vaccination 	        & $3.6$ 	&  
\textit{``I've got the second shot of vaccine! I need to buy sport drink when I go home...and most important of all, food for the cat!''}  \\ 
&	&		\\	
{\bf Theme 2: Breaking news} 	& {\bf $21.3$}      &   	\\
Clinical trial and use authorization 	& $8.0$ 	&  
\textit{``Approval by the Ministry of Health, Labour and Welfare (MHLW) of the only vaccine for new coronavirus from US pharmaceutical giant Pfizer.''} 	\\
Effectiveness of vaccination 	& $7.5$ 	& 
\textit{``It is reported that mRNA vaccines are effective against corona \#corona \#mRNA vaccine \#effective.''}   \\
Booking vaccination appointment  &  $5.8$	& 
\textit{``The unprecedented scale of Mass vaccination: What is the preparation status of local municipalities?''}   \\
  &	    &		\\
  {\bf Theme 3: Politics}    & 	{\bf $17.2$} 	 &		\\
Opinion about politics 	                    &  $9.6$ 	 &  
\textit{``To the idiots in the government: if you can inoculate corona vaccine to all Japanese citizens, you can hold Olympic and Paralympic, but if you can't, cancel them.''}  		\\
Opinion about mass media 	                &  $4.2$ 	 &  
\textit{``The mass media raised fears with coronas, and now they are raising fears with vaccines. Media should report the facts unbiasedly instead of raising fears.''}    \\
Vaccination policy 		                    & $3.4$ 	& 
\textit{``It may be better to take a wait-and-see approach to the vaccination. It could be bad.''}   \\
  &	        &	    \\
  {\bf Theme 4: Conspiracy, Humour}    & 	{\bf $11.8$} 	 &	  \\
Population control	&  $4.2$ 	& 
\textit{``They developed the corona vaccine for the purpose of the Deep State agenda: global human enslavement, depopulation and money making!''} \\
Effect on the body 		&  $3.1$ 	&   \textit{``It's very exciting to be able to connect to 5G when you are vaccinated!''} \\
Internet meme 	                                        & $4.5$ 	& \textit{``Vac-vac-cine-cine! Vac-vac-cine-cine! vacvac! cinecine! cinecinevacvac!''} \\        
\hline
\end{tabular}
\end{table*}

\begin{figure}[t]
  \centering
       \includegraphics[width=1.0\linewidth]{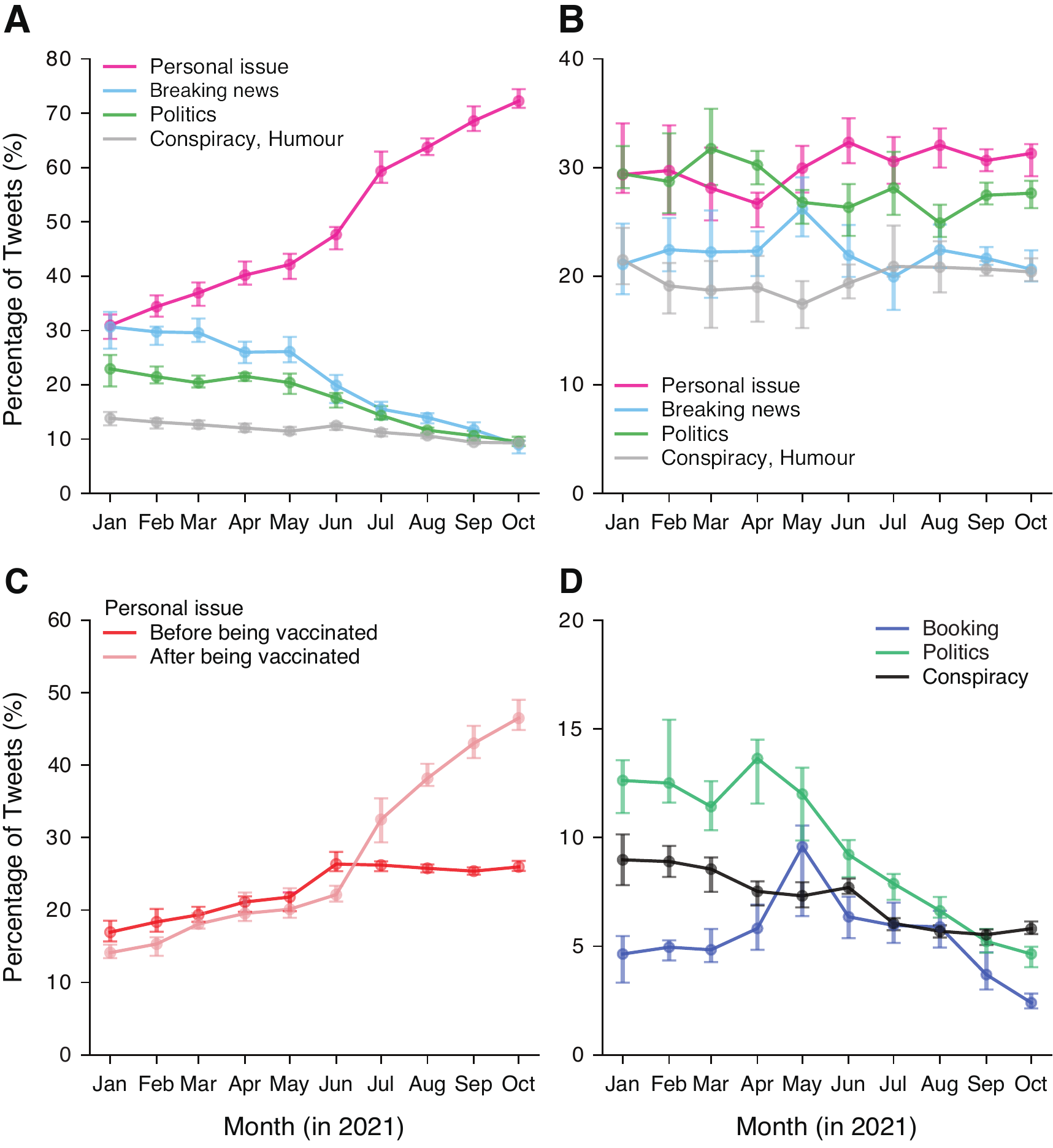}
  \caption{ {\bf Popularity of the themes or topics.} 
  Each line represents the percentage of tweets in each theme or topics over time. The percentage is calculated monthly considering (A), (C), and (D) a sample of the vaccine-related tweets (1 million tweets: sample 2), and (B) only frequently retweeted tweets (i.e.\ retweeted more than 10 times in a day).}
  \label{fig:Trend_theme}
\end{figure}

\subsection{Evolution of the Popularity of Themes}

Previous research has shown that the number of tweets about a particular topic reflects the users' attention to that topic~\cite{Kwak2010,Lehmann2012}. We thus estimated the popularity of tweets for each topic (grouped in 4 major themes, see previous section) to monitor temporal changes in the interest of users (Fig.~\ref{fig:Trend_theme}A). Personal issue (Theme 1) continuously increased, starting at nearly $30\%$ to over $70\%$ by the end of the study period. Breaking news (Theme 2) and Politics (Theme 3), on the other hand, declined steadily respectively from nearly $30\%$ and $25\%$ to less than $10\%$, dropping more significantly after June, when vaccination became available for people under 65 years old (the majority of Twitter users). Conspiracy and humour (Theme 4) also reduced slightly in the period and overall remained relatively low. 
We further validated this analysis by creating a subset of keywords for each theme (SI Appendix, Table S3) and then extracting all tweets of each theme from the original data set (24 million tweets) (SI Appendix, Fig. S1A). 
The linear regression analysis (SI Appendix, Fig. S1B and Table S4) showed a statistically significant increase in the tweets about Personal issue (theme 1) and a decrease in the other themes, with Breaking news (theme 2) and Politics (Theme 3) decreasing five times in comparison to Conspiracy and humour (Theme 4). These trends reveal a shift on the concerns of Tweeter users, who initially shared their attention over personal issue (individual aspect), collecting information from the news (knowledge acquisition), and government decisions (the course of the vaccination campaign), towards focusing mostly on personal issue once the vaccination campaign was effectively implemented on the general population.

The evolution of specific topics reflects finer aspects of the opinion dynamics. The combined topics about personal issues before being vaccinated (i.e.\ personal view and personal schedule) increased after May followed by a slight decrease after August (Fig.~\ref{fig:Trend_theme}C). This pattern reflects increasing concerns with the vaccination and the Tokyo Olympic Games that ended in early August. The combined topics about user's experience after being vaccinated (i.e.\ live reports, journal, perception, and preparation) showed a sharp increase after June, when the vaccination of the general population began (Fig.~\ref{fig:Trend_theme}C). In contrast, conspiracy theories (population control and effect on the body) decreased steadily indicating that education built up confidence in the vaccines (Fig.~\ref{fig:Trend_theme}D). Opinions about the booking of vaccination appointments peaked in May, when the booking system was launched. Opinions about politics peaked in April and then decreased substantially reflecting an initial criticism towards the government for the late implementation of the mass vaccination, followed by approval once the campaign rolled out. Again, we validated these findings by extracting the corresponding tweets using a sub-set of keywords for each topic or aggregated topic (SI Appendix, Table S3) and confirmed the trends (SI Appendix, Fig. S2), with a low prevalence of words related to conspiracy theories ($5.8 \%$). This result also confirms that the initial concerns about the government and reliability on the vaccines became secondary once the vaccination reached most of the population, and personal experiences became dominant.

\subsection{Shift in Interest at Critical Events}
Specific events may have social and individual consequences and affect the public opinion and discussion of different themes. Four critical events marked the vaccination campaign in Japan during 2021, the various stages of the vaccination campaign and the Tokyo Olympic Games (Fig.~\ref{fig:Twitter_TS}). To test our hypothesis of critical events on the opinion dynamics, we performed an interrupted time series regression~\cite{Bernal2017} to estimate the changes in the popularity of themes (SI Appendix, text for detail). In this analysis, the level parameter indicates a shift in (the relative) attention whereas the slope indicates the rate of popularity of a given theme. Table~\ref{tab_ITS} shows that Politics (theme 3) was the most affected theme by these events. The impact of the start of the vaccination of the general population and the Tokyo Olympic Games on the public opinion was larger than that of the other critical events, and they affected all aspects of public opinion. The vaccination of health workers positively shifted and accelerated the popularity of the politics theme, likely because of increasing expectations on rolling out mass vaccination. The vaccination of the elderly population only positively shifted the trend. On the other hand, both the vaccination of the general population and the Tokyo Olympic Games negatively shifted the interest on politics, suggesting relative less concerns with government policies. Furthermore, the start of the vaccination of the general population increased the rate of tweets about practical advice, personal experience, and news about the reliability of the vaccine. Finally, the start of the Tokyo Olympic Games caused a shock of interest in personal issues that remained nearly constant afterwards, likely because the large vaccination coverage achieved during this period (SI Appendix Fig. S3).

\subsection{Spread of Opinions}
A tweet is a unidirectional process of sharing information with the community. Retweeting, on the other hand, is a social process where users engage and share tweets to spread opinions on their own social network~\cite{Boyd2010}. The analysis of 75 984 321 retweets by 3 917 181 users (sample 3) showed a higher prevalence of retweets about Personal issue (theme 1) and Politics (theme 3) in comparison to Breaking news (theme 3), and Conspiracy and humour (theme 4) (Fig.~\ref{fig:Trend_theme}B). Those observations align with the theory of complex contagion since users mostly engaged with tweets (by retweeting) related to personal experiences and political opinion rather than tweets sharing hard-to-verify information, such as vaccine reliability and conspiracy theories, that might have negative consequences and affect the credibility of the user retweeting~\cite{State2015}. Similar to the popularity of certain topics, the social process is also intensified at certain periods (SI Appendix, Fig. S4). 
For instance, the topic of booking an appointment exhibits a sharp peak in May, coinciding with the popularity of this topic, whereas the topic of politics declined after April, when the vaccination of the elderly population started.

\begin{table*}[htbp] \centering
\caption{Changes in the popularity of each theme at critical events, i.e.\ the start of each stage of the vaccination campaign and the Tokyo Olympic Games. Statistically significant changes ($\text{p-value} < .05$) are highlighted.
\label{tab_ITS}} 
\vspace{3mm}
\begin{tabular}{p{4cm}p{3cm}p{3cm}p{3cm}p{3cm}}
\hline 
&  \textbf{Theme 1}   &  \textbf{Theme 2}  &  \textbf{Theme 3} &  \textbf{Theme 4}  \\ 
\hline  
\textbf{Health workers} \\
\hspace{3mm}Level &      1.21 &     -3.93 &      \textbf{6.74}  &  -0.14 \\
\hspace{3mm}95\% C.I. &      [-4.51,6.93] &     [-8.88,1.03] & [1.27,12.2]  &  [-1.45,1.18] \\
\hspace{3mm}Slope &     -0.01 &    -0.20 &      \textbf{0.55} & -0.05 \\
\hspace{3mm}95\% C.I. & [-0.32,0.31] & [-0.55,0.15] & [0.30,0.80] & [-0.14,0.03] \\
\hline
\textbf{Elderly population} \\
\hspace{3mm}Level &     -0.71 &    -3.25 &      \textbf{5.59} &     -0.98 \\
\hspace{3mm}95\% C.I. & [-3.54,2.13] & [-7.79,1.28] & [1.60,9.58] & [-2.11,0.15] \\
\hspace{3mm}Slope &     -0.14 &      0.13 &      0.06 &      0.07 \\
\hspace{3mm}95\% C.I. & [-0.28,0.01] & [-0.17,0.42] & [-0.15,0.27] & [-0.01,0.14] \\
\hline
\textbf{General population} \\
\hspace{3mm}Level &      0.53 &      0.23 &     \textbf{-2.19} &      0.045 \\
\hspace{3mm}95\% C.I. & [-1.61,2.66] & [-2.39,2.85] & [-3.80,-0.59] & [-0.76,0.85] \\
\hspace{3mm}Slope & \textbf{0.33} & \textbf{0.34} & 0.07 & \textbf{-0.08} \\
\hspace{3mm}95\% C.I. & [0.21,0.46] & [0.14,0.54] & [-0.02,0.15] & [-0.13,-0.04] \\
\hline
\textbf{Olympic Games} \\ 
\hspace{3mm}Level & \textbf{4.57} & 0.54 &     \textbf{-2.17} & \textbf{-0.89} \\
\hspace{3mm}95\% C.I. & [2.80,6.35] &  [-1.42,2.51] & [-3.19,-1.15] & [-1.32,-0.46] \\
\hspace{3mm}Slope &     \textbf{-0.27} &      \textbf{0.18} &      0.02 & \textbf{0.03} \\
\hspace{3mm}95\% C.I. & [-0.37,-0.17] & [ 0.04,0.32] &     [-0.05,0.08] & [0.00,0.07] \\
\hline  
\end{tabular}
\end{table*}

\section{Discussion}
The first year of the COVID-19 pandemics was marked by a rush to control spread and develop efficient vaccines. Although most high-income countries pushed to launch mass vaccination as early as possible, the government of Japan was criticized for not reacting timely, particularly given public concerns with the Tokyo Olympic Games that had been postponed to the summer of 2021. Part of the delay was associated with fears that the Japanese population could resist vaccination, given past experiences with HPV vaccines~\cite{Hanley2015}. Twitter provides a platform to monitor in real-time the public debate and engagement in topics of relevance to health and policy, and not least social and economic implications of government decisions~\cite{Kwak2010}. We leveraged the textual information on tweets and performed a topic analysis of 114 357 691 vaccine-related tweets to identify 15 topics further grouped into four major themes: 1) Personal issue, 2) Breaking news, 3) Politics, and 4) Conspiracy and humour during the vaccination campaign in Japan. We found a major shift in public interest, with users splitting their attention to various themes early in the campaign and then focusing on personal issues, as trust in vaccines and policies built up with an effective vaccination campaign.
Increased trust helped to reduce the prevalence of tweets about conspiracy and humour, which have negative impact on vaccine uptake~\cite{Loomba2021}. Previous research using social media (Twitter and Reddit) to study the public perception of COVID-19 vaccination in different countries were limited in sample size and did not cover the whole vaccination campaigns. Therefore, only topics related to breaking news~\cite{Wu2021,Lyu2021,Kwok2021}, and politics~\cite{Lyu2021} were identified. 
We show however that personal issues 
are common topics and fundamental for an effective campaign due to social support in times of uncertainty. The interrupted time series regression analysis showed that the start of vaccination of the general population and the Tokyo Olympic Games affected public opinion more than other critical events. Public opinion on politics was the most significantly affected debate, positively shifting the attention early in the vaccination campaign and negatively later. 
In addition, a social dialogue was maintained with tweets about personal issue mostly retweeted when the vaccination reached the adult population, that is the most active user group in Twitter.

The online data set is a convenience sample and thus the study population is limited to those using Twitter in Japan. 
To minimize potential sampling biases, we re-sampled the original data of 6 million users to remove temporal effects. Unlike standard survey studies, we were unable to collect socio-demographic information and thus could not stratify the analysis to age-group, location, or education and gender~\cite{Solis2021,Nomura2021}. Stratification would help us to assess the extent that certain social groups (e.g.\ adults vs. elderly) and locations (e.g.\ Tokyo during the Olympic Games) were affected. 
Furthermore, the inclusion criteria of tweets with the keyword ``vaccine'' may have captured tweets not relevant to COVID-19, as for example those tweets related to HPV vaccine or pet vaccination. To assess this aspect, we manually reviewed tweets and found that most of them were not contaminated by discussion of other types of vaccines. Our methodology allowed to cluster and monitor the evolution of public opinion that moved from an exploratory phase of gathering information and criticizing the government to social support by sharing personal experiences once confidence in the vaccination was established and practical issues became relevant.

\section{Methods}
\subsection{Data Collection}
We downloaded all the tweets written in Japanese including the word ``waku-cine'' (vaccine in Japanese) posted between January 1, 2021 and October 31, 2021. The data set was provided by the NTT DATA Corporation (\url{https://www.nttdata.com}). The study period was chosen to include few weeks before the launch of the vaccination campaign in Japan (February 17, 2021) until few weeks after the end of the Tokyo Olympic Games when the vaccination rate reached $75\%$ of the Japanese population (October 17, 2021). The data set contains 114 357 691 tweets. We extracted all the tweets except ``quote tweets'' or ``mentions'' to other tweets. We further collected data of the tweet text, the time stamp (posting time), and whether the tweet was original or a retweet. Using data from Our World in Data (\url{https://ourworldindata.org}), we obtained the daily incidence (number of new cases) of COVID-19 and the vaccination rate (the percentage of the population who received at least one dose) of COVID-19 vaccine in Japan~\cite{Mathieu2021}.

\subsection{Data Processing}
The data processing and analysis were performed using Python software, version 3.9.7 (Python Software Foundation). We first manually identified and removed all the tweets posted by bots, then extracted the plain text from the remaining tweets and removed Emojis. Afterwards, we segmented each text into Japanese words using the morphological analyzer MeCab~\cite{MeCab} and removed stop words that have little analytic value, e.g.\ ``kore'', ``sore'', ``suru'', meaning respectively ``this'', ``it'', and ``do'' in Japanese. Finally, we changed words to their root forms, e.g.\ ``boku'' to ``watasi'' (``I'' in Japanese) or ``Utta'' to ``Utsu'' (``inject'' in Japanese). This normalization corresponds to e.g.\ ``viruses'' to ``virus'' or ``went'' to ``go'' in English.

\subsection{Topic Modeling} 
The statistical Latent Dirichlet Allocation (LDA) model~\cite{Blei2003} implemented in the {\it Gensim} Python package~\cite{Rehurek2010} was used to identify topics in the Twitter data. 
Before the topic modeling analysis, we removed rare words, i.e.\ words appearing in less than 1 000 tweets that corresponds to $0.0004\%$ of the tweets, and the most frequent words ``waku-cine'' (vaccine) and ``sessyu'' (vaccination). 
To determine the number of topics, we trained LDA models with different numbers of topics and maximized the topic coherence score~\cite{Roder2015} that is a robust measure of how topics are meaningful (i.e.\ interpretable) to humans.

\section*{ACKNOWLEDGMENTS}
This research was conducted as part of “COVID-19 AI \& Simulation Project” run by Mitsubishi Research Institute commissioned by Cabinet Secretariat.
The methodology used in this study was developed through the support from  
JSPS KAKENHI (Grant Numbers JP18K11560, JP19H0113, JP21H03559, JP21H04571, JP22H03695, JP22K12285 and JPJSBP120202201), 
JST CREST (Grant Number JPMJCR1401), 
JST PRESTO (Grant Number JPMJPR1925), 
and AMED (Grant Number JP21wm0525004).

\section*{Author Contributions}
Y.N., Y.T., T.S., and R.K. are co-first authors. 
R.K. conceived the study and R.K and L.R designed the study. 
T.H., M.T., N.Y, and M.K. collected and preprocessed the data.
Y.N., Y.T., T.S., T.U., and R.K. analyzed data.
Y.N., Y.T., and R.K. created figures.
R.K. and L.R wrote the manuscript. 
All authors discussed the results and contributed to finalize the manuscript. 
R.K. supervised the project.

\bibliography{ms}

\end{document}


\clearpage   
\widetext
\begin{center}
\textbf{\large Supplemental Materials: Evolution of the public opinion on COVID-19 vaccination in Japan}
\end{center}
\setcounter{equation}{0}
\setcounter{figure}{0}
\setcounter{table}{0}
\setcounter{page}{1}
\makeatletter
\renewcommand{\theequation}{S\arabic{equation}}
\renewcommand{\thefigure}{S\arabic{figure}}
\renewcommand{\thetable}{S\arabic{table}}
\renewcommand{\bibnumfmt}[1]{[S#1]}
\renewcommand{\citenumfont}[1]{S#1}

\section*{Interrupted time series regression Analysis}
We used an interrupted time series regression~\cite{SI_Bernal2017} to quantify the impact of major events (e.g.,\ the start of the Olympic games, see Fig. 1) on the popularity of a theme in tweets. 
The statsmodels Python package~\cite{SI_Seabold2010} was used for this analysis. 
The theme was defined based on a subset of keywords (Table S3) and the analysis is based on all the tweet data set (sample 1: 24 million tweets). 
We assumed the following regression model: 
\begin{align}
 Y_t = \beta_0 + \beta_1 T + \beta_2 X_t + \beta_3 T X_t + u_t,
\end{align}
where $Y_t$ is the percentage of a theme in tweets at time $t$ (day), $T$ is the number of days from the start of the observation period ($T=0$) (i.e., 30 days before each event), $X_t$ is a dummy variable that equals to 0 (1) before (after) the event, respectively, and $u_t$ is the error term. 
Here, $\beta_0$ represents the baseline of the popularity (in percentage) at $T=0$, $\beta_1$ represents the slope before the event, and $\beta_2$ and $\beta_3$ represent the level and slope change after the event, respectively. 

Time series often exhibit autocorrelation, that is, the error terms are correlated over time, whereas the regression analysis assumes that the error terms $u_t$ are uncorrelated. 
To evaluate the confidence intervals of the estimated parameters (Table 2), we calculate Newey–West standard error~\cite{SI_Newey1987,SI_Andrews1991}, also known as the Heteroskedasticity- and Autocorrelation- Consistent (HAC) standard error, that are robust to the autocorrelation. 
We also calculate Newey–West standard error to evaluate the confidence intervals of the linear regression analysis (Table S4). 
\clearpage

\begin{figure*}
\centering
\includegraphics[width=\textwidth]{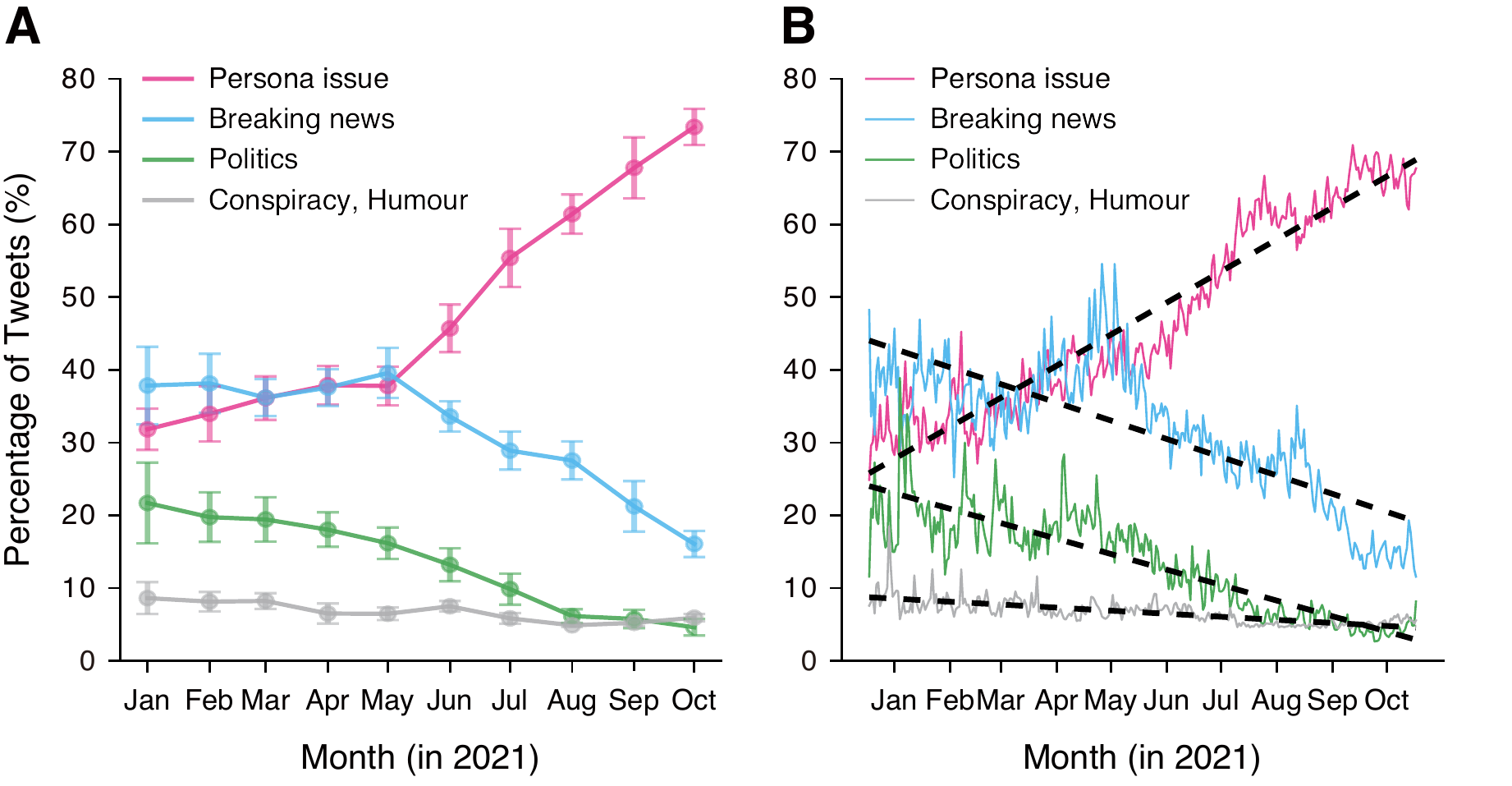}
\caption{Popularity of the themes defined based on the subset of keywords (Table S3). 
Each line represents the percentage of tweets of each theme over time. 
The percentage was calculated for each month (A) and day (B).
Dashed lines in panel (B) represent the fitted lines obtained by the linear regression analysis. }
\end{figure*}


\begin{figure*}
\centering
\includegraphics[width=\textwidth]{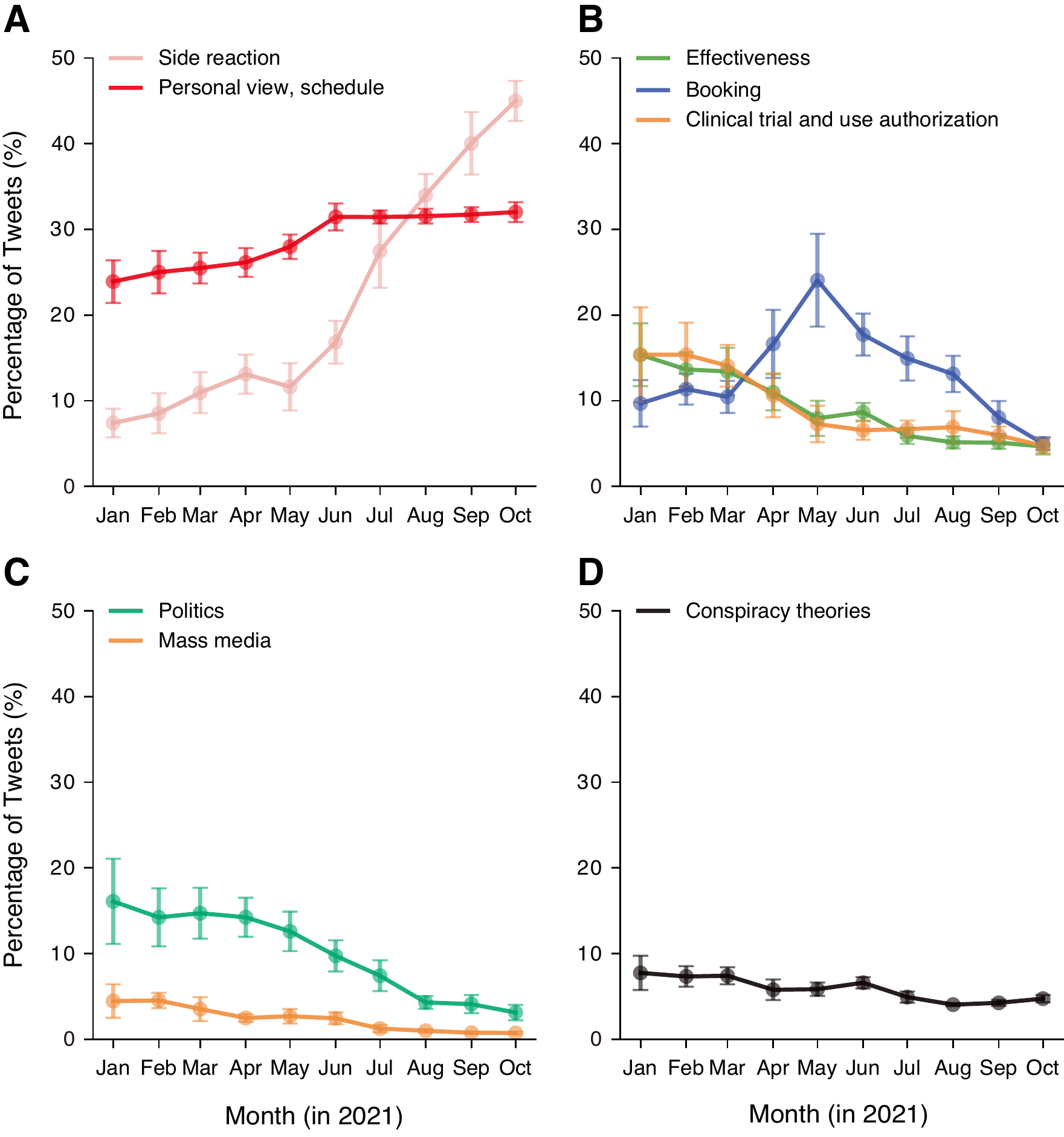}
\caption{Popularity of the sub-themes defined based on the subset of keywords (SI Appendix, Table S3). 
 (A) Sub-themes related to Personal issue that corresponds to Theme 1 (Table 1) identified by the topic modeling (Fig. 3C). 
 (B) Sub-themes related to Breaking news that corresponds to Theme 2 (Fig. 3D). 
 (C) Sub-themes related to Politics that corresponds to Theme 3 (Fig. 3D). 
 (D) Sub-themes related to Conspiracy theories corresponding to Theme 4 (Fig. 3D). 
 }
\end{figure*}

\begin{figure*}
\centering
\includegraphics[width=\textwidth]{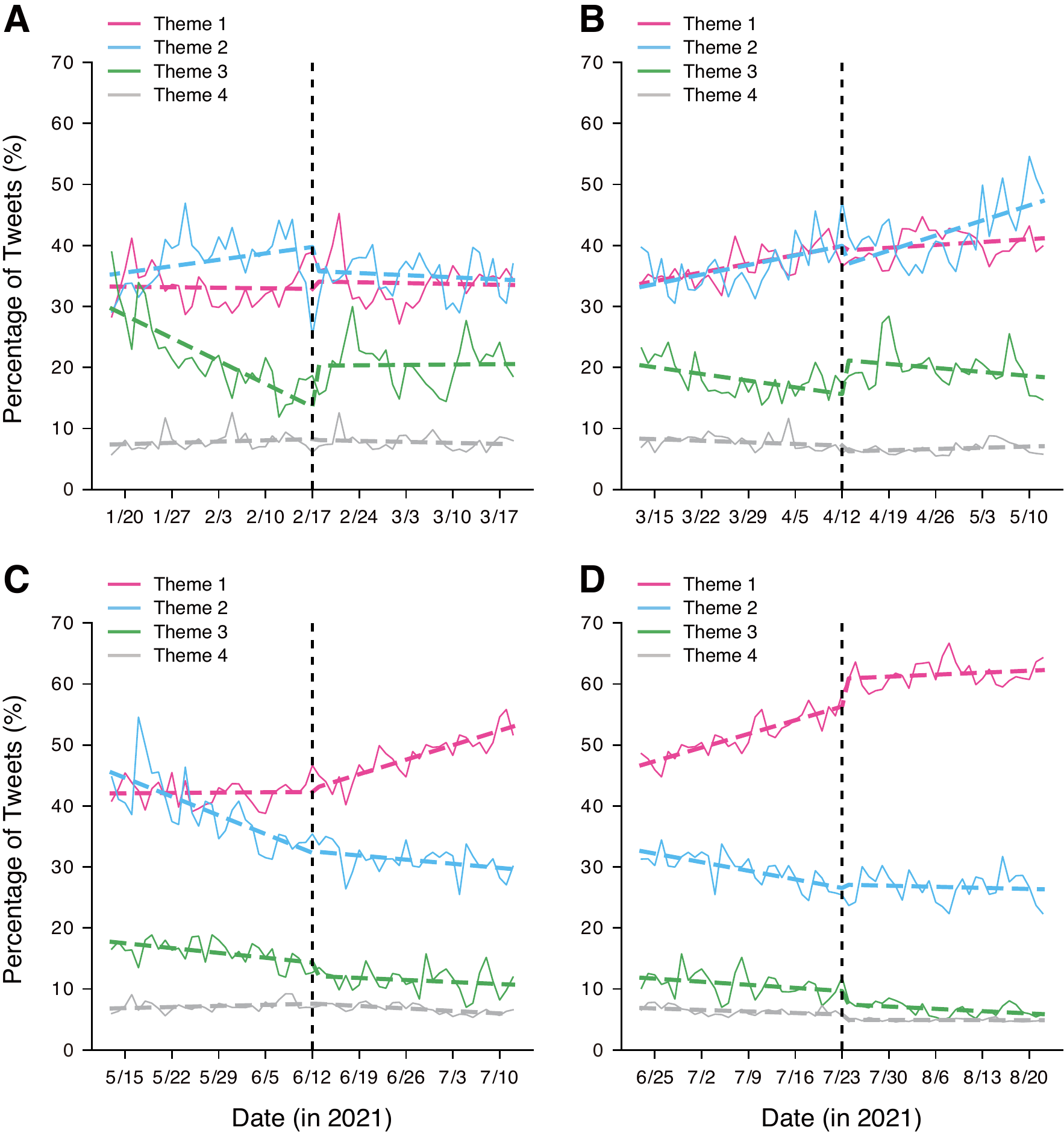}
\caption{Impact of social events on the popularity of the themes.  
We first calculate the popularity (i.e., the percentage of tweets) of four themes defined by subsets of keywords (Table S3), theme 1: Personal issue (magenta), theme 2: Breaking news (cyan), theme 3: Politics (green), and theme 4: Conspiracy, Humour. 
We then applied the interrupted time series analysis to the time series of the popularity of each themes for four major events during the vaccination period: 
(A) Vaccination start for health workers, 
(B) Vaccination start for older people, 
(C) Vaccination start for general population (under 65), and 
(D) Start of Olympic games in Tokyo. 
}
\end{figure*}

\begin{figure*}
\centering
\includegraphics[width=\textwidth]{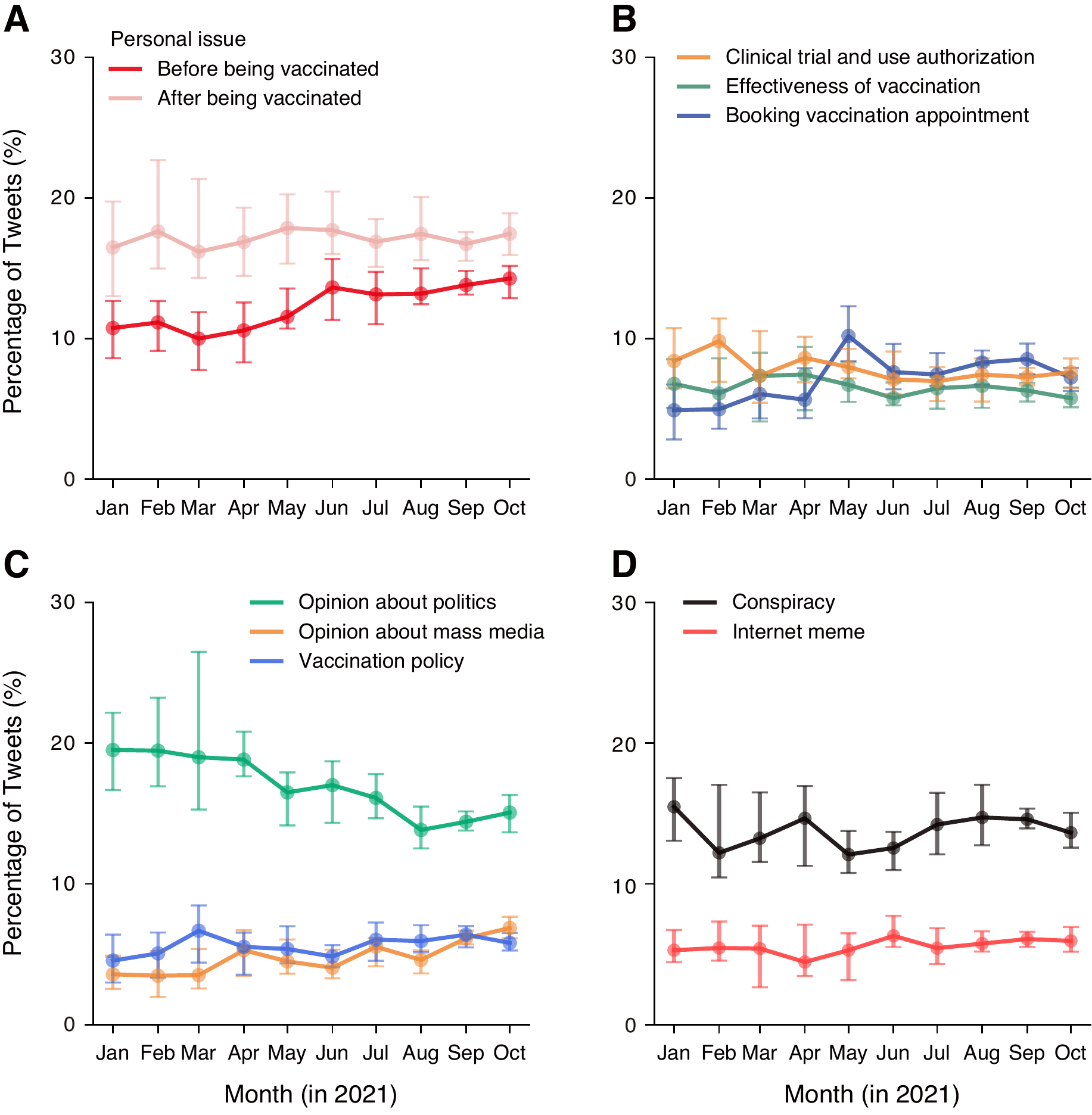}
\caption{Popularity of identified topics based on the topic model (Table 1, Table S2) in frequently retweeted tweets. 
(A) Theme 1: Personal issue, (B) Theme 2: Breaking news, (C) Theme 3: Politics, and (D) Theme 4: Conspiracy, Humour. 
Note that the topics in theme 1 (A) and 4 (D) are aggregated as in Fig. 3A and D. The topics of Before (After) being vaccinated represent a combined topics of personal view and personal schedule (live reports, journal, perception, and preparation). The topics of Conspiracy represent a combined topics of population control and effect on the body. 
}
\end{figure*}

\begin{table*}[tbhp]
\centering
\caption{Top 50 used words in vaccine-related tweets: a million tweets (sample 2).} 
\vspace{3mm}	
\begin{tabular}{llr}
\hline 
Rank & Word  & No. of tweets \\
\hline \hline   
1 & Covid-19 & 242627 \\
2 & get ( vaccinated) & 174956 \\
3 & 2nd & 112841 \\
4 & 1st & 60387 \\
5 & side reaction & 58640 \\
6 & news & 55742 \\
7 & say & 53600 \\
8 & I & 52211 \\
9 & reservation & 51621 \\
10 & think & 47654 \\
11 & today & 47372 \\
12 & run (a fever) & 46781 \\
13 & Japan & 44660 \\
14 & infection & 35938 \\
15 & arm & 35610 \\
16 & Olympics & 31095 \\
17 & see & 30786 \\
18 & tomorrow & 30517 \\
19 & Pfizer & 30395 \\
20 & finish & 26087 \\
21 & side effect & 25587 \\
22 & heat & 25571 \\
23 & hurt & 25032 \\
24 & elderly & 25002 \\
25 & Yahoo & 24760 \\
26 & do & 23903 \\
27 & can & 23202 \\
28 & go & 23076 \\
29 & come & 22235 \\
30 & earlier & 22078 \\
31 & effect & 21964 \\
32 & myself & 21876 \\
33 & injection & 21442 \\
34 & yesterday & 20609 \\
35 & make & 20385 \\
36 & ask & 19914 \\
37 & can get (vaccinated) & 18700 \\
38 & mask & 18634 \\
39 & consider & 18460 \\
40 & scary & 18270 \\
41 & exist & 18252 \\
42 & work & 18169 \\
43 & healthcare workers & 18036 \\
44 & start & 17903 \\
45 & dose & 17538 \\
46 & pain & 17231 \\
47 & government & 17140 \\
48 & death & 17004 \\
49 & cine-cine & 16816 \\
50 & die & 16469 \\\hline  
\end{tabular}
\end{table*}

\begin{table*}[tbhp]
\centering
\caption{Top contributing words for each topic identified by the topic model (LDA). }
\vspace{3mm}		
\tabcolsep = 10pt
\begin{tabular}{p{5.5cm}p{2cm}p{6.5cm} }
 \hline
 Theme and topic	&   No. of tweets ($\%$)  &   Top terms contributing to the topic model  \\
 \hline  \hline   
{\bf Theme 1: Personal issue}       & {\bf $49.8$}      &   	\\
Personal view on vaccination        & $17.2$ 	    &	I, think, myself, scary, absolutely, feeling, alright      \\  
Personal schedule of vaccination    & $5.8$ 	    &	today, tomorrow, finish, clinic, schedule, this week, next week\\
Live reports of before/after vaccination        & $3.2$ 	&	go back, venue, pain, swell, 30 minutes, sleepy, wait\\
Journal about vaccination experience        & $13.4$ 	&	2nd time, side effect, 1st time, yesterday, job, fine, temperature\\
Perception after vaccination        & $6.6$ 	    &	sore, feel, arm, injection, left arm, incredibly, discomfort\\
Preparation for vaccination	        & $3.6$ 	    & fever, condition, 2nd day, prepare, lighten, better, helpful        \\
    &       &      \\
{\bf Theme 2: Breaking news} 	& {\bf $21.3$}      &   	\\
Clinical trial and use authorization    & $8.0$ 	&	Phizer, Moderna, "Ministry of Health, Labour and Welfare", development, start, clinical trial, approval        \\
Effectiveness of vaccination        & $7.5$ 	    &	effect, mRNA, report, death, research, Moderna, Isreal       \\
Booking vaccination appointment	    &  $5.8$        &   booking, preperation, group, campaign, possible, system, municipality  \\
     &      &      \\
{\bf Theme 3: Politics}     & 	{\bf $17.2$} 	 &		\\ 
Opinion about politics	            &  $9.6$ 	        & Japan, measures, nation, government, impossible, declaration of a state of emergency, Tokyo\\
Opinion about mass media            &  $4.2$ 	        &	anxiety, information, press, rumor, mass media, explication, fact   \\
Vaccination policy                  &  $3.4$ 	        & 3rd time, recent, terrible, tweet, video, shit, laugh                     \\
     &      &      \\
{\bf Theme 4: Conspiracy, Humour}    & 	{\bf $11.8$} 	 &	  \\
Population control          &  $4.2$ 	    & 	population, human being, world, cause, Conspiracy theory, reduction\\
Effect on the body          &  $3.1$ 	    &	children, 5G, freedom, destroy, discrimination\\
Internet meme               &  $4.5$ 	    &	cine-cine  \\
\hline   
\end{tabular}
\end{table*}

\begin{table*}[ht]	\centering
\caption{Lists of keywords for the four themes and their sub-themes. Keywords are manually assigned from the result of the topic modeling and reading typical tweets in each topic. }    
\vspace{3mm}		
\tabcolsep = 10pt
\begin{tabular}{p{5.5cm}p{8.5cm}}
 \hline
 Theme and topic	&   Keywords   \\
 \hline\hline
{\bf Theme 1: Personal issue} 	       &   	\\
Personal view, schedule	   &    
absolutely, consider, feel, feeling, husband, I, impossible, inject, mother, myself,next month, next week, normal, normal, okay, okay, parent, quick, quickly, safe, scared, schedule, think, this week, today, tomorrow, understand, weird, you  \\
Side reaction  	   &      
1st time, 2nd time, 37°C, 38°C, 39°C, ache, adverse reaction, antipyretic, calonal, chill, fever, headache, high fever, joint pain, lassitude, loxonin, malaise, migraine, muscle pain, nausea, normal temperature, pain, painful, painkiller, physical condition, pyrexia, shiver, side effect, side reaction, slight fever, temperature, tiredness, vomit  \\
&			\\	
{\bf Theme 2: Breaking news} 	&   	\\
Clinical trial and authorization 	&  
approval, AstraZeneca, clinical trial, development, manufacturing, MHLW, Moderna, Pfizer, pharmaceutical Company, supply, US
 	\\
Effectiveness 	    &  
dead,death, effect, effective, effectiveness, examination, expert, influence, Israel, mRNA, mutant strain, mutation, study
 	\\
Booking             & 
acceptance, access, availability, booking, booking site, center, connect, elderly, family doctor, group, large-scale, log in, occupied, phone, priority, quota, reception, self-defense forces, system, venue, workplace 
 	\\
&	   		\\
{\bf Theme 3: Politics}   &		\\
Politics	          &  
administration,  Bach, benefits, compensation, councilor, Dr. Omi, election, games, Go To Travel campaign, 
government,  governor, IOC, IOC president, Kishida prime minister, LDP (Liberal Democratic Party), lockdown, mayor of Akashi city, minister, Ministry of Defense, Mr. Abe,Mr. Kono, Mr. Nishimura,  Mr. Norihisa Tamura, Ms. Junko Mihara, Ms. Marukawa, Ms. Yuriko Koike, official residence, olympic, olympic minister, opening ceremony, opposition party, overwhelmed health care system, paralympic, policy, politician, presidency, prime minister, quasi-state of emergency,  
regulatory reform, special measures Law, state of emergency declaration, subcommittee, torch relay,
vaccine passport, without spectators, zero answer
  		\\
Mass media	          &     
agitating, audience, biased reporting, gag, incitement, mass communication, mass media, media hype, news, press, press attitude, report, TV Asahi  \\
&	  	    \\
{\bf Theme 4: Conspiracy, Humour}    &		\\
Conspiracy theories	    &       
5G, autism, Bill Gates, brainwashing, connection, conspiracy theories, cytokine storm,destruction, DNA, electromagnetic wave,  evil, gene, human experimentation, humankind, hydrogen, magnet, microchip, money, population, population reduction, prophecy, reduction, Rothschild, Russia, warning, zombie         \\
\hline
\end{tabular}
\end{table*}
\clearpage

\begin{table*}\centering
\caption{Linear regression analysis of the popularity of the theme (Fig. S1). 15 topics identified by the topic modeling were grouped into four themes (Table 1 and S2): Theme 1: Personal issue, Theme 2: Breaking news, Theme 3: Politics, and Theme 4: Conspiracy and humour. }
\vspace{5mm} 
\begin{tabular}{lcccc}
 \hline  
& \textbf{Theme 1} & \textbf{Theme 2} & \textbf{Theme 3} & \textbf{Theme 4} \\
 \hline  
\hspace{3mm}    Intercept   & 25.8 &  44.0  & 24.0 & 8.75 \\ 
\hspace{3mm}    $95\%$ C.I. & [23.9,27.6] & [41.2,46.8] & [22.2,25.8] & [8.22,9.27] \\  
%
\hspace{3mm}    Slope        &  0.142  & -0.082  & -0.070   & -0.014 \\
\hspace{3mm}    $95\%$ C.I.  & [0.133, 0.152]  & [-0.097,-0.067]  & [-0.078, -0.061] & [-0.017, -0.011] \\  
\hline  
\end{tabular}
\end{table*}




\section*{Reference}
\bibliography{supplement}